\documentstyle[12pt]{article}
\title{\bf The  top quark mass and flavor mixing \\in a Seesaw model of 
        Quark Masses}
\author{\vspace*{1em}\\
        {\bf T. Morozumi}\thanks{E-mail address :
         morozumi@theo.phys.sci.hiroshima-u.ac.jp}
        ~and ~{\bf T. Satou}\thanks{E-mail address : 
t\_{}satou@kakuri2\_{}pc.phys.sci.hiroshima-u.ac.jp }\\
        Department of Physics, Hiroshima University\\
        1-3-1 Kagamiyama, 739 Higashi Hiroshima, Japan\\
         \\
        {\bf M. N. Rebelo}\thanks{E-mail address : rebelo@beta.ist.utl.pt}\\
        CFIF/IST and Departamento de F\'{\i}sica\\
        Instituto Superior T\'ecnico\\
        Av. Rovisco Pais, 
        P-1096 Lisboa Codex, Portugal\\
        \\
       {\bf M. Tanimoto}\thanks{E-mail address :
tanimoto@edserv.ed.
ehime-u.ac.jp}\\ 
       Science Education Laboratory, Ehime University\\                  
       Bunkyo-cho, 790 Matsuyama, Japan\\
       }
\date{}
\begin{document}
\setlength{\baselineskip}{14pt}
\renewcommand{\thesubsection}{\arabic{subsection}}
\maketitle
\begin{picture}(0,0)
       \put(335,390){HUPD-9704}
       \put(335,375){FISIST/6-97/CFIF}
       \put(335,360){hep-ph/97xxxxx}
\end{picture}
\vspace{-24pt}
\thispagestyle{empty}
%%%%%%%%%%%%%%% abstract %%%%%%%%%%%%%%%%%%%%%%%%%
\begin{abstract}
The top quark mass and the flavor mixing are studied  in the context
of a Seesaw model of Quark Masses based on the
gauge group $SU(2)_L \times  SU(2)_R \times U(1)$.
Six isosinglet quarks are introduced  to give rise
to the mass hierarchy of ordinary quarks.
In this scheme, we reexamine a mechanism for the
generation of the top quark mass.
It is shown that, in order to prevent the Seesaw mechanism
to act for the top quark, the mass parameter of its
isosinglet partner  must be much smaller
than the breaking scale of $SU(2)_R$. As a result the fourth
lightest up quark must have a mass of the order of 
the breaking scale of $SU(2)_R$,  and a large mixing
between the right-handed top quark and its singlet partner occurs.
We also show that this  
mechanism is compatible with the mass
spectrum of light quarks and their flavor mixing.
\end{abstract}
%%%%%%%%%%%%%%%%% Intro %%%%%%%%%%%%%%%%%%%%%%%%
\section{\bf Introduction}
\def\valpha{{ \mbox{\boldmath$\alpha$}}}
\def\vbeta{{ \mbox{\boldmath$\beta$}}}
\def\vecg{{ \mbox{\boldmath$y$}}}
\def\vecu{{ \mbox{\boldmath$U$}}}
\def\veca{{ \mbox{\boldmath$a$}}}
\def\vecv{{ \mbox{\boldmath$V$}}}
\def\vecu2{{ \mbox{\boldmath$u$}}}
\def\vecv2{{ \mbox{\boldmath$v$}}}
%%%%%%%%%%%%%%%%%%%% 1 %%%%%%%%%%%%%%%%%%%%%%%%%%%%%%%%%
The Seesaw mechanism \cite{Yana}, \cite{Grs} was initially invented to
explain the smallness of neutrino masses. Also, in a different framework,
the smallness of masses of quarks other than the top quark
compared to the scale of the electroweak symmetry breaking
can be explained by a Seesaw Model 
of Quark Masses \cite{Bere}, \cite{ChMo}, \cite{Raj}, 
\cite{Wali}, \cite{Moha}.
  In the context of $SU(2)_L \times SU(2)_R \times U(1)
$ models including isosinglet quarks, it has been shown that the light 
ordinary quark masses are proportional to 
 the breaking scales of $SU(2)_L$,  ($\eta_L$) and $SU(2)_R$,
($\eta_R$) and are 
 inversely proportional to the isosinglet 
quark mass $M$, i.e., $O\left({ {\eta_L \eta_R} \over M } 
\right)$.
 Conventionally, the isosinglet quark mass $M$ is assumed to 
be much larger than  $\eta_R$.
  This assumption leads to 
an explanation for the smallness of quark masses compared to the 
scale of the electroweak symmetry breaking. 
Though the Seesaw mechanism explains the smallness of the
mass of the five flavors 
from the up quark to the bottom quark, it has not been shown 
that the top quark with a mass of 
$ O(\eta_L)$ can be incorporated into the same scheme.
In this letter, we study  the top quark mass as well as
the mass hierarchy of the up and down quark sectors 
in the context of the Seesaw mechanism.
We show  that the mass formulae for the light quarks proposed
before is not valid for the top quark 
and  must be replaced by a new one.
The mass hierarchy of the light quarks and flavor mixing 
are studied in the same context by ref. \cite{Raj},
\cite{Wali}, \cite{Moha} under the assumption that
the isosinglet quark diagonal mass parameter is much larger than the
breaking scale of $SU(2)_R$. 
However, under the same assumption the top quark mass
would also be much smaller than the breaking scale of
$SU(2)_L$ unless the corresponding Yukawa coupling between
isosinglet and isodoublet quarks is chosen to be very large.
 As we show later, if  the diagonal mass
parameter for the isosinglet partner of the top quark is 
much smaller than  $\eta_R$,
the Seesaw mechanism does not act for the top quark
and its mass can be  kept at the scale of $\eta_L$ without introducing
a large Yukawa coupling. In this case the heavier mass eigenstate
is as light as $\eta_R$ rather than being given by the singlet mass parameter.
Because one of the mass eigenstates is as light as $\eta_R$,
the flavor mixing and the stability of the light quark masses
against the inclusion of flavor off-diagonal Yukawa couplings is a 
non-trivial problem, we study the light quark mass
spectrum and flavor mixing taking into consideration  
the special r\^ole played by the top quark.   
We obtain  the approximate mass
formulae for quark masses and show that the mass hierarchy 
is stable against  flavor mixing.\\
 
This paper is organized as follows. In section 2, we present the 
results for the diagonalization of the mass matrix for the top quark and its
isosinglet partner. The mixing between singlet and
doublet quarks is discussed  both for left- and right-handed
chiralities.
In section 3, the mass spectrum of both light and heavy
quarks is obtained by introducing flavor
off-diagonal couplings.
In section 4, the mixing angles are obtained.
In section 5, we summarize the results.
%%%%%%%%%%%%%%%%%%%% Sec.2 %%%%%%%%%%%%%%%%%%%%%%%%%%%%
\section{\bf The top quark mass and singlet-doublet mixing in a Seesaw model}

The gauge group of the model is $SU(2)_L \times SU(2)_R \times U(1)$
 and the ordinary quarks,  isosinglet quarks and the relevant Higgs scalars
are assigned to the following gauge group representations:
%%%%%%%%%%% Fermion %%%%%%%%%%%%%%
\def\VL{V_L}
\def\VLD{{V_L}^\dagger}
\def\VR{V_R}
\def\VRD{{V_R}^\dagger}
\begin{eqnarray}
\psi_L^i= \left( \begin{array}{c}   u^i\\
                               d^i \end{array} \right)_L
&:& (2,1,1/3),\nonumber\\
\psi_R^i=\left( \begin{array}{c}      u^i\\
                              d^i \end{array} \right)_R
&:&(1,2,1/3),\nonumber\\
U^i_{L,R}&:&(1,1,4/3),\nonumber\\
D^i_{L,R}&:&(1,1,-2/3),\nonumber\\
\phi_L&:&(2,1,1),\nonumber\\
\phi_R&:&(1,2,1),
\end{eqnarray}
where $i=1,2,3$.
%%%%%%%%%%%%%
By introducing an isodoublet Higgs $\phi_L$ ($\phi_R)$ for
$SU(2)_L$ ($SU(2)_R)$, the
Yukawa interaction between doublet and singlet quarks 
and the mass term for the singlet quarks of the model are given by
\begin{eqnarray}
{\cal L}_y&=&-y^{ij}_{LD}\bar{\psi^i_L}\phi_L
D^j_R-y^{ij}_{LU}\bar{\psi^i_L}\tilde{\phi_L}U^j_R\nonumber\\
&&-y^{ij}_{RD}\bar{\psi^i_R}\phi_R
D^j_L-y^{ij}_{RU}\bar{\psi^i_R}\tilde{\phi_R}U^j_L+ (h.c.)\nonumber\\
&&-\bar{U^i}M^i_U U^i-\bar{D^i}M^i_D D^i,\nonumber\\
\label{eqn:lag}
\end{eqnarray}
where $i$ and $j$ are summed over from 1 to 3;
$y_{L\{R\}D(U)}$ is the strength of the Yukawa coupling between the down(up) type
left-handed \{right-handed\} isodoublet quark and isosinglet
quark;
$y_{L(R)}$ are $3\times 3$ matrices; $M_{U}^i(M_D^i)$ are given 
by $M_U^1=M_U,\, M_U^2=M_C, \, M_U^3=M_T, \, M_D^1=M_D, \, M_D^2=M_S$
and $M_D^3=M_B$.
Without loss of generality,
the singlet quark mass matrix  can be transformed into a real 
diagonal matrix through a  
bi-unitary transformation.
The scale of the singlet mass parameter is going to be set 
later. 

We first focus on the case in which the flavor mixing is absent
and study the top quark sector.
The mass eigenstates and eigenvalues for the top quark and its 
isosinglet partner
are obtained by diagonalizing the two by two matrix: 
\def\ubar{\bar u}
\def\cbar{\bar c}
\def\tbar{\bar t}
\def\Tbar{\bar T}
%%%%%%%%%%%%%%%  Heavy quark t %%%%%%%%%%%%%%%%%%%%%%%%%%%
\begin{eqnarray}
&{}&\left( \bar{t} ~ \bar{T} \right)_L
\left( \begin{array}{cc} 0& y_L \eta_L \\
                         y_R^{\ast}\eta_R& M_T \end{array} \right)
\left( \begin{array}{c} t\\
                        T \end{array} \right)_R\nonumber\\
&=&\left( \bar{t^m} ~ \bar{T^m} \right)_L
\left( \begin{array}{cc} m_t&0 \\
                         0& m_T \end{array} \right)
\left( \begin{array}{c} t^m\\
                        T^m \end{array} \right)_R.
\label{eqno:tres}                        
\end{eqnarray}
%%%%%%%%%%% t mass %%%%%%%%%%%%%%%%%%%%%%%%%%%%%%%%%%
The mass eigenvalues and the mixing angles for the left-handed 
chiralities  are obtained by
diagonalizing   $MM^\dagger$ :
\begin{eqnarray}
MM^\dagger=
\left( \begin{array}{cc} |y_L|^2  {\eta_L}^2  &  y_L  \eta_L M_T\\
                 y_L^{\ast} \eta_L 
M_T & {M_T}^2 + {|y_R|}^2 {\eta_R}^2 \end{array} \right),
\end{eqnarray}
where $\eta_{L(R)}$ are the vacuum expectation values of the neutral 
Higgs particles,
\begin{eqnarray}
\langle \phi_{L} \rangle=\left( \begin{array}{c}0\\
                        \eta_L\end{array} \right),\nonumber\\
\langle \phi_{R} \rangle=\left( \begin{array}{c}0\\
                        \eta_R\end{array} \right).
\end{eqnarray}
The vacuum expectation values are related to 
the masses of the charged $SU(2)_{L(R)}$ gauge bosons by
\begin{eqnarray}
M_L^2&=&{1\over 2}g^2\eta_L^2,\nonumber\\
M_R^2&=&{1 \over 2}g^2\eta_R^2,
\end{eqnarray}
where $g$ is the $SU(2)_{L(R)}$ gauge coupling constant.
Therefore
the mass eigenvalues are written in terms of the gauge boson masses as:
\begin{eqnarray}
 & m_t& \cong \sqrt{2}\left( \frac{|y_L y_R| }{g^2} \right) M_L
\frac{1}{ \sqrt {\left(\frac{ |y_R| }{g} \right)^2 + 
\left( \frac{M_T}{\sqrt{2}M_R} \right)^2 } },\label{mt}\\ 
 & m_T&  \cong \sqrt{2}M_R \sqrt{\left(\frac{ |y_R| }{g} \right)^2 + 
\left( \frac{M_T}{\sqrt{2}M_R} \right)^2}.
\label{eqn:mass}
\end{eqnarray}
%%%%%%%%%%%%%%%%%
If  $M_T \gg {M_R}$, the lighter mass eigenvalue $m_t$ in 
Eq.(\ref{mt}) is 
reduced to the well known Seesaw formulae for light quarks, i.e., 
  $O\left( \frac{M_L M_R}{M_T} \frac{|y_L y_R| }
{g^2} \right)$.  However, for the top
quark, we should take the limit $M_T \ll M_{R}$.
 Then the suppression factor $ {M_R \over M_T} $
is now absent  and its mass is given by
\begin{equation}
m_t \cong \sqrt{2} {|y_L| \over g} M_L,
\label{eqn:topmass}
\end{equation}
which is at the
scale of $M_{L}$. On the other hand, the heaviest eigenvalue $m_T$ in
Eq.(\ref{eqn:mass}) is
\begin{equation}
m_T \cong \sqrt{2} {|y_R| \over g} M_R,
\label{eqn:capt}
\end{equation}
which  is at the symmetry breaking scale of $SU(2)_R$
instead 
of being given by
the singlet mass parameter $M_T$. 
To reproduce the value of the top 
quark mass in Eq.(\ref{eqn:topmass}),  the Yukawa coupling $y_L$ must be a few
times larger than the gauge coupling $g$. We assume that it is
still within the perturbative region.
There is a simple  reason why the top quark mass is
determined  by $M_{L}$
 and its partner's mass is proportional  to  $M_{R}$.
To show this, let us ignore the mass term  of the singlet
quark 
and keep only singlet-doublet mixing terms.
 Then the Yukawa term  of the top
quark and its singlet partner is
\begin{equation}
{\cal L}= - y_L \eta_L \bar{t_L} T_R  - y_R^{\ast} \eta_R
\bar{T_L} t_R + (h.c.).  
\label{eqn:sd}
\end{equation}
{}From  Eq.(\ref{eqn:sd})
 the mass term can be diagonalized
by the following rotations,
\begin{eqnarray} 
\left( \begin{array}{c} t^m\\
                               T^m \end{array} \right)_L &=&
\left( \begin{array}{c} t\\
                               T \end{array} \right)_L,\\
\left( \begin{array}{c} t^m\\
                               T^m \end{array} \right)_R &=&
\left( \begin{array}{c} T\\
                               t\end{array} \right)_R.
\end{eqnarray}
Then  the mass  for  $t^m$ is 
$|y_L|  \eta_L$ and the mass for $T^m$ is 
$ |y_R| \eta_R$.
We note that the isosinglet and doublet mixing is maximal 
for  the right-handed sector.
The large mixing between the top  quark and its singlet partner 
predicted in this model should be possible to observe 
in an experiment of $ t \bar t$ production.

In the limit where flavor mixing is absent
we obtain, from the Lagrangean of Eq.(\ref{eqn:lag}),
mass matrices for the pairs (u,U), (c,C), (d,D), (s,S)
and (b,B), generically denoted by (q,Q), similar to the one for the
pairs (t,T) in Eq.(\ref{eqno:tres}). The eigenvalues are also given by 
expressions of the form Eq.(\ref{mt}) and Eq.(\ref{eqn:mass}) with $M_T$
replaced by the corresponding mass parameter $M_Q$. In order
for the Seesaw mechanism to act, so that the masses we obtain for
the light quarks are much smaller than $M_L$, we need
$M_Q \gg M_R$. Then Eq.(\ref{eqn:topmass}) and 
Eq.(\ref{eqn:capt}) are replaced by 
\begin{eqnarray}
&m_q& \cong 2 \frac{|y_Ly_R|}{g^2} M_L \left (\frac{M_R}{M_Q} \right ),\\
&m_Q& \cong M_Q.
\end{eqnarray}
If $M_R$ is fixed, we can estimate the order of magnitude of $M_Q$
for each case. In these cases the mixing is suppressed by a 
factor of  $({\eta_L
\over M_Q}) $ for the left-handed components and $( {\eta_R
\over M_Q})$ for the  right-handed ones \cite{Moha}. 

In the next section we extend our analysis to the case where
flavor mixing mass terms are present.

%%%%%%%%%%%%%%%%%%%%%%%%%%%%%%%%%%%%%%%%
\section{\bf Mass formulae including flavor mixing  in the Seesaw model}
In this section we derive the mass formulae in the case with
mixings among different quark flavors. 
We start with  the six by six mass matrix
%%%%%%%%%% Def M %%%%%%%%%%%%%%%%%%%%%%%%
\begin{equation}
M=\left( \begin{array}{cc}0&\eta_L y_L\\
                        \eta_R y_R^\dagger&M_{Diag} 
           \end{array} 
    \right),
\end{equation}
where $M_{Diag}$ is a three by three diagonal mass matrix,
\def\MD{M_{Diag}} 
%%%%%%%%%%%%% Def M_D %%%%%%%%%%%%%%%%%
\begin{equation}
\MD=\left[ \begin{array}{ccc}M_U&\ &\ \\
                             \ &M_C&\ \\
                             \ &\ &M_T \end{array}
    \right],
\end{equation}
and $y_L$ and $y_R$ are rank 3 matrices.
The eigenvalue equation for the quark masses is given by 
%%%

\begin{equation}
\det(MM^\dagger-\Lambda)=\det
\left[ \begin{array}{cc}\eta^2_L y_L y_L^\dagger-\Lambda&y_L M_{Diag} \eta_L \\
                        \eta_L M_{Diag} y_L^\dagger&\eta_R^2
y_R^\dagger y_R + 
\MD^2-\Lambda \end{array}
\right]=0.
\label{eig.eq}
\end{equation}
The equation  which determines the eigenvalues  of order of $
{\eta_L}^2$
(or smaller than  ${\eta_L}^2$)
is reduced to a cubic equation:
%%%%
\begin{equation}
\det\left[  y_L \left\{1-\MD
(\eta_R^2 {y_R}^\dagger y_R + {\MD}^2)^{-1} \MD
\right\}y_L^\dagger - \lambda \right]=0,
\label{cub}
\end{equation}
where we use the normalized eigenvalue $ \lambda=
 {\Lambda \over \eta_L^2}$.
This equation determines the ordinary quark masses.  
%%%
We further use the expansion as follows,
\begin{equation}
(\eta_R^2 y_R^\dagger y_R +\MD^2)^{-1}={1\over M_0}\left[
1-({1\over M_0}\Delta M^2{1\over M_0})+({1\over
M_0}\Delta M^2{1\over M_0})^2+\cdots\right]{1\over M_0},
\label{exp}
\end{equation}
%%%%%%%%%%%%%% Def M_0 %%%%%%%%%%%%%%%%%%
\begin{eqnarray*}
%%%
M_0&=&\left[ \begin{array}{ccc}M_U&\ &\ \\
                              \ &M_C&\ \\
                              \ &\ &\eta_R\sqrt{(y_R^\dagger y_R)_{33}}
 \end{array}
     \right], \\
%%%
\Delta M^2&=&\eta_R^2y_R^\dagger y_R+\MD^2-M_0^2,\\
\end{eqnarray*}
where we assume $M_U,M_C \gg \eta_R \gg M_T$.
As a result the unperturbed
part of $(\eta_R^2 y_R^\dagger y_R +\MD^2)$ is given by $M_0^2$
rather than by $\MD^2$. 
%%%%%%%%%%%%%%%%%%%%%%%%%%%%%%%%
 Eq.(\ref{cub}) is expressed in the simple form,  
\begin{equation}
\det \left[ y_L  Y  y_L^\dagger-\lambda 
     \right]=0,\label{eig.eq2}
\end{equation}
where the leading terms of $Y$, obtained by using the
expansion of Eq.(\ref{exp}), are as follows:
%%%%%%%%%%%%%%%%%%% Def Y_elements %%%%%%%%%%%%%%%
\begin{eqnarray}
Y_{11}&=
&X_U^2[(y_R^\dagger y_R)_{11}-\frac{(y_R^\dagger y_R)_{13}
(y_R^\dagger y_R
)_{31}}{(y_R^\dagger y_R)_{33}}],\nonumber\\
Y_{12}&=&X_UX_C[(y_R^\dagger y_R)_{12}-\frac{(y_R^\dagger
y_R)_{13}(y_R^\dagger y_R)_{32}}{(y_R^\dagger y_R)_{33}}],\nonumber\\
Y_{13}&=&
\frac{X_UX_T(y_R^\dagger y_R)_{13}}{(y_R^\dagger y_R)_{33}},\nonumber\\
Y_{22}&=&X_C^2[(y_R^\dagger y_R)_{22}-\frac{(y_R^\dagger
y_R)_{23}(y_R^\dagger y_R)_{32}}{(y_R^\dagger y_R)_{33}}],\nonumber\\
Y_{23}&=&
\frac{X_CX_T(y_R^\dagger y_R)_{23}}{(y_R^\dagger y_R)_{33}},\nonumber\\
Y_{33}&=&1-\frac{X_T^2}{(y_R^\dagger
       y_R)_{33}}+\frac{X_T^4}
{(y_R^\dagger y_R)_{33}},
\end{eqnarray}
where $Y^\dagger=Y$ and $X_U,X_C$ and $X_T$ are  given by
%%%%%%%%%%%%%%%%%%%%%%%%% Def X's %%%%%%%%%%%%%%%%%
\begin{eqnarray}
X_U&=&{\eta_R \over M_U},\nonumber\\
X_C&=&{\eta_R \over M_C},\nonumber \\
X_T&=&{M_T\over{\eta_R\sqrt{(y_R^\dagger y_R)_{33}}}}.
\end{eqnarray}
%%%%%%%%%%%%%%%% Eig.eq %%%%%%%%%%%%%%%%%%
The eigenvalue equation now becomes
\begin{equation}
F(\lambda)=-\lambda^3+\lambda^2 Tr[y_L^\dagger y_L Y]-\gamma
\lambda+|\det y_L|^2\det Y=0.\label{eig.eq3}
\end{equation}
%%%%%%%%%%%%%% Approx Eig.eq %%%%%%%%%
The coefficients for each term of
Eq.(\ref{eig.eq3}) are
\begin{eqnarray}
Tr[y_L^\dagger y_L Y]&\cong& (y_L^\dagger y_L)_{33},\nonumber\\
\gamma&\cong& X_C^2|\valpha_{R2}|^2 
|\valpha_{L2}|^2(y_L^\dagger y_L)_{33},\nonumber\\
\det Y &\cong& X_U^2 
X_C^2\left(|\valpha_{R1}|^2
|\valpha_{R2}|^2-|\valpha_{R1}^\dagger\cdot\valpha_{R2}|^2\right),
\label{albe}
\end{eqnarray}
where we are keeping the leading term for each coefficient.
Here $\valpha$ and $\vbeta$ are 
complex vectors which are related to the three complex vectors
 $ (\vecg_{1},\vecg_{2},\vecg_{3})$    of  the matrix $y_L$ and 
$y_R$.  To show the relation, let us first write $y_L$ and 
$y_R$ as follows,
\begin{eqnarray}
y_L&=&\left( \vecg_{L1},\vecg_{L2},\vecg_{L3} \right),\nonumber\\
y_R&=&\left( \vecg_{R1},\vecg_{R2},\vecg_{R3} \right).
\label{eqn:y}
\end{eqnarray}
Then  $\valpha$ and $\vbeta$ are defined by 
\begin{eqnarray}
\valpha_{LI}&=&\vecg_{LI}-\frac{\vecg_{L3}^\dagger 
\cdot\vecg_{LI}}{|\vecg_{L3}|^2}\vecg_{L3},\nonumber\\
\valpha_{RI}&=&\vecg_{RI}-\frac{\vecg_{R3}^\dagger
\cdot\vecg_{RI}}{|\vecg_{R3}|^2}\vecg_{R3},\nonumber\\
\vbeta_{L1}&=&\valpha_{L1}-\frac{\valpha_{L2}^\dagger \cdot
\valpha_{L1}}{|\valpha_{L2}|^2}\valpha_{L2},\nonumber\\
\vbeta_{R1}&=&\valpha_{R1}-\frac{\valpha_{R2}^\dagger \cdot
\valpha_{R1}}{|\valpha_{R2}|^2}\valpha_{R2},
\label{eqn:vec}
\end{eqnarray}
where $I=1,2$.
Geometrically $\valpha_{I}(I=1,2)$ are the projections of $\vecg_{I}$
onto  the plane perpendicular to the vector $\vecg_{3}$, and 
$\vbeta_1 $ is the projection of $\valpha_{1}$ onto the line
perpendicular to $\valpha_{2}$
 and $\vecg_{3}$.
To find the solutions of Eq.(\ref{eig.eq3}), we first set
the order of magnitude of
 $X_U $ and $X_C$ as follows.
If  we neglect flavor mixing, 
in the Seesaw model, the up quark mass is given by
 $  {M_L M_R
\over M_U} $ with  $ y_{L(R)} \simeq g$. Therefore the
Seesaw suppression factor $X_U={M_R \over M_U}$ is of the
order of ${m_u \over M_L} \simeq 10^{-5}$. With a similar
argument, we have 
$X_C={M_R \over M_C} \simeq 10^{-2}$ by assuming that
the strength of the Yukawa coupling does not depend 
significantly on the flavor.
By setting the magnitude of $X_U$ and $X_C$ in this way, we obtain an
approximate mass formula for the case of flavor mixing.
The point is that in the approximation that the strengths 
of all Yukawa couplings do not depend significantly on the flavor,
the eigenvalue equation Eq.(\ref{eig.eq3}) has the following form,
\begin{equation}
F(\lambda)=-\lambda^3+ O(1)\lambda^2 - O( X_C^2)
\lambda+O(X_U^2 X_C^2)=0.\label{eig2.eq3}
\end{equation} 
The solutions of Eq.(\ref{eig2.eq3}) are hierarchical, i.e., 
$\lambda= O(X_U^2), \, O(X_C^2),\, O(1)$
with $ X_U \ll X_C \ll 1$.
Taking into account the structure of the Yukawa coupling strengths 
the solutions are of the form:
%%%%%%%%%%%%%%% def aplpha beta %%%%%%%%%%%%%%%%
%
\begin{eqnarray}
m_t^2&\cong& |\vecg_{LU3}|^2\eta_L^2,\nonumber\\
m_c^2&\cong& X_C^2 |\valpha_{RU2}|^2
|\valpha_{LU2}|^2\eta_L^2,\nonumber\\ 
m_u^2&\cong& X_U^2|\vbeta_{RU1}|^2|\vbeta_{LU1}|^2\eta_L^2
.
\label{eqn:up}
\end{eqnarray}
The heavier eigenvalues are also obtained as follows,
\begin{eqnarray}
m_U^2&\cong& M_U^2,\nonumber\\ 
m_C^2&\cong& M_C^2,\nonumber\\
m_T^2&\cong& |\vecg_{RU3}|^2\eta_R^2.
\label{eqn:hup}
\end{eqnarray}
%%%%%%%%%%%%%%%%%% masses %%%%%%%%%%%%%%%%%%%%%
In the case of the down type quarks the masses are obtained by assuming 
$M_D \gg M_S \gg M_B \gg \eta_R$:
\begin{eqnarray}
m_b^2&\cong& X_B^2 |\vecg_{LD3}|^2|\vecg_{RD3}|^2 \eta_L^2,\nonumber\\
m_s^2&\cong& X_S^2
|\valpha_{RD2}|^2 |\valpha_{LD2}|^2 \eta_L^2,\nonumber\\ 
m_d^2&\cong&
X_D^2 |\vbeta_{RD1}|^2|\vbeta_{LD1}|^2  \eta_L^2,\nonumber\\
m_D^2&\cong& M_D^2,\nonumber\\
m_S^2&\cong& M_S^2,\nonumber\\
m_B^2&\cong& M_B^2.
\label{down}
\end{eqnarray}
with $X_B=
{\eta_R \over M_B}$,\, $X_S={\eta_R \over M_S}$ and $X_D={\eta_R
\over M_D}$.

We would like to stress the following points.
\begin{itemize}
\item The mass formulae for the five light quarks are of the 
Seesaw type.  They are proportional to
$\eta_L, \, \eta_R$ and inversely proportional to the singlet
quark mass. They are also proportional to
the appropriate projection
of the Yukawa couplings $y_{LUI}, y_{RUI}(I=1,2)$ for the up and the 
charm quarks
and   $y_{LDI}, y_{RDI}(I=1,2,3)$ for the down type quarks.
 \item The top quark mass ($m_t$) is propotional to $\eta_L$ and
the length of the vector $\vecg_{LU3}$. The other mass eigenvalue ($m_T$) is
propotional to $\eta_R$ and the length of $\vecg_{RU3}$.
 \item The masses of the remaining five heavier quarks are  
 given to a good approximation by the mass parameters
$M_{Ui} \, (i=1,2)$ and $M_{Di} \, (i=1,2,3)$.
\end{itemize}
\section{\bf Flavor mixing }
\hspace*{1em}
In order to find the mixing angles (CKM matrix) among left (right)
chiralities of quarks, it is convenient to perform
a unitary transformation among ordinary 
quarks such that the singlet-doublet Yukawa couplings,
 i.e., $y_L$ and $y_R$  become triangular
matrices,
\def\yuUp{y11'}
\def\yuCp{y12'}
\def\yuTp{y13'}
\def\ycTp{y23'}
\def\ycUp{y11'}
\def\ycCp{y22'}
\def\ytUp{y31'}
\def\ytCp{y32'}
\def\ytTp{y33'}
\begin{equation}
y_{L\{R\}U(D)}
  = U_{U(D) L \{R\}}
 \left[ \begin{array}{ccc} \times  &0  &0 \\
                             \times  & \times &0 \\
                             \times   & \times   & \times  \end{array}
    \right],
\end{equation}
where the $U$'s are unitary matrices. This decomposition can always be 
done as proved in the Appendix.
Further we introduce a new basis denoted by $u',d'$ which is 
related to the original basis by the unitary matrices,
\begin{eqnarray}
u'_L&=&U_{UL}^\dagger u_L,\nonumber \\
u'_R&=&U_{UR}^\dagger u_R,\nonumber \\
d'_L&=&U_{DL}^\dagger d_L,\nonumber \\
u'_R&=&U_{DR}^\dagger d_R.
\end{eqnarray}

In this new basis, the $6 \times 6$ matrix $M$
is transformed into
\begin{equation}
M'= \left[ \begin{array}{cc} U_L^\dagger & 0 \\
                               0         & 1
            \end{array}
            \right]
     M 
  \left[ \begin{array}{cc} U_R & 0 \\
                               0         & 1
            \end{array}
            \right].
\end{equation}

The explicit form of $M'$ for the up type quarks is given by

\begin{equation}
M' =  \left[ \begin{array}
{cccccc}  0  & 0  &0  & \beta_{LU11} \eta_L & 0 & 0 \\
          0  & 0  &0  & \beta_{LU21} \eta_L& |\mbox{\boldmath
$\alpha$}_{LU2}|
 \eta_L & 0 \\
          0  & 0  &0  & \alpha_{LU31}\eta_L  &
\alpha_{LU32}\eta_L &  | \mbox{\boldmath $y$}_{LU3}| \eta_L\\
         \beta_{RU11}^* \eta_R& \beta_{RU21}^* \eta_R
  &\alpha_{RU31}^*  \eta_R & M_U & 0  &  0\\
          0  & | \mbox{\boldmath $\alpha$}_{RU2}| \eta_R 
& \alpha_{RU32}^* \eta_R
&  0  & M_C  &  0\\
          0  & 0   & |\mbox{\boldmath $y$}_{RU3}| \eta_R
&  0  &  0  &   M_T  \end{array}
    \right],
\end{equation}
where $|\beta_{LU11}|=|\vbeta_{LU1}|$ and $|\beta_{RU11}|=|\vbeta_{RU1}|$;
their phases can be obtained from the Appendix;
$\valpha_{LD2}$, $\vecg_{LD3}$,
$\valpha_{LU2}$ and $\vecg_{LU3}$ are defined by Eq.(\ref{eqn:y}) 
and Eq.(\ref{eqn:vec}).
In this form, $u'$ only couples to $U$. Therefore the
up quark mass is determined by the mass of the heaviest singlet
quark $U$ and is  not affected by the presence of the other
lighter singlet quarks.
The $c'$ couples to both $ U$ and $C$. However its mass is mainly 
determined by the mass of $C$ and the effect of $U$ is tiny
because 
 $M_U \gg M_C$.   
As a result the matrix $M'$ can be approximated  by the following block
diagonal matrix $M'^0$, 
\begin{equation}
M'^0 = \left[ \begin{array}
{cccccc}  0  & 0  &0  & \beta_{LU11} \eta_L & 0 & 0 \\
          0  & 0  &0  & 0& |
\valpha_{LU2}
| \eta_L & 0 \\
          0  & 0  &0  & 0  &
 0 &| \vecg_{LU3}| \eta_L\\
         \beta_{RU11}^* \eta_R& 0
  &     0& M_U & 0  &  0\\
          0  & | \valpha_{RU2}| \eta_R &
 0
&  0  & M_C  &  0\\
          0  & 0   &| \vecg_{RU3}|
 \eta_R&  0  &  0  &   M_T \end{array}
    \right].
\end{equation}
The off-diagonal elements of $M'$ which do not appear in
$M'^0$
can be
treated as a perturbation.
In this approximation, each doublet quark only couples to one
of the singlet quarks. So the eigenvalues and eigenstates are
obtained by  diagonalizing two by two matrices.
It is interesting to see that the eigenvalues of $M'^0$
agree with those obtained in Eq.(\ref{eqn:up}) and Eq.(\ref{eqn:hup}) 
with the same assumption for the
singlet quark mass parameters.
In addition the mass eigenstates are defined by the following rotation
between the doublet quark $q'$ and the singlet quark $Q$,
\begin{equation} 
        \left( \begin{array}{c}
                q^m\\ 
                Q^m
        \end{array} \right) =
        \left( 
        \begin{array}{cc}
                \cos \theta_q & - \sin \theta_q e^{-i \phi_q} \\
                \sin \theta_q e^{i \phi_q} & \cos \theta_q 
        \end{array} \right) 
        \left(\begin{array}{c}
                q' \\ 
                Q
        \end{array} \right).
\end{equation}
\begin{table}[t]
        \begin{center}
        \begin{tabular}{|l|l|l|} \hline
        Doublet-Singlet  & $\tan\theta_{qL} \exp( i \phi_{qL}) 
$  &     $ \tan\theta_{qR} \exp(i \phi_{qR})$\\ 
        \hline \hline
        $ u' - U $  & $ {\beta}_{LU11} \eta_L / M_U$
                & 
                   $ {\beta}_{RU11} \eta_R/ M_U$ 
        \\ \hline
$   c' - C      $ & $|{\valpha}_{LU2}| \eta_L /M_C$
                & $  |{\valpha}_{RU2}| \eta_R /M_C $ 
                \\ \hline
$  t' - T $ 
                &  $|\vecg_{LU3}| \eta_L M_T /|\vecg_{RU3}|^2 \eta_R^2$
                & $|\vecg_{RU3}| \eta_R / M_T      $
        \\ \hline 
$   d' - D      $ & $ {\beta}_{LD11}\eta_L /M_D$
                & $ {\beta}_{RD11}\eta_R /M_D $ 
                \\ \hline

$   s' - S      $ & $ |{\valpha}_{LD2}|\eta_L/M_S$
                & $  |{\valpha}_{RD2}|\eta_R /M_S $ 
                \\ \hline

$   b' - B      $ & $ |\vecg_{LD3}| \eta_L/M_B$
                & $  |\vecg_{RD3}|\eta_R /M_B $ \\
                \hline
        \end{tabular}
        \caption{Singlet and Doublet mixing angles}
\label{tab:bound}
        \end{center}
\end{table}
In Table \ref{tab:bound}, we show the mixing angles 
between singlet and doublet quarks
for both chiralities.
It can be seen that the singlet to doublet mixing for the five
light quarks is strongly suppressed being at least of the order
$\eta_R (\eta_L) \over M$ for right (left) -handed quarks.
For the top quark, there is a suppression of the left-handed 
mixing angles of the order $ {\eta_L M_T \over \eta_R^2}$ while
the right-handed mixing angle is not suppressed.
In the following analysis, we only keep the right-handed
mixing angle for the top quark and its singlet partner and  
set the other singlet to doublet mixing angles 
to zero.
In this approximation, the left-handed charged currents and the CKM
matrix are given by

\begin{equation}
{j_{\mu L}^-} ={\overline {{u^m}_{Li}} } 
\gamma_\mu V_{Lij} {d^m}_{Lj} ,
\end{equation}
where the CKM matrix is written, in terms of the unitary matrices
defined before, as:
\begin{equation}
V_L= {U_{UL}}^\dagger  {U_{DL}}.
\end{equation}
Within this approximation, we do not have FCNC for the left-handed
neutral isospin current.
On the other hand, the right-handed charged currents are given by
\begin{eqnarray}
{j_{\mu R}^-} &=& {\overline {{u^m}_R} } 
\gamma_\mu V_{R 1j} {d^m}_{Rj} 
            + {\overline {{c^m}_R} }
 \gamma_\mu V_{R 2j} {d^m}_{Rj}\nonumber \\ 
            &+& \cos \theta_{tR} 
 {\overline {{t^m}_R} } \gamma_\mu 
                       V_{ R 3j} {d^m}_{Rj}
                                   +     \sin \theta_{tR} 
 {\overline  {{T^m}_R} } \gamma_\mu  
                 V_{R 3j} {d^m}_{Rj},
\end{eqnarray}
where $V_R= {U_{UR}}^\dagger  U_{DR}$
and $\theta_{tR}=\tan^{-1}( |y_{RU3}| \eta_R / M_T) $
as shown in Table 1.
The right-handed neutral isospin currents  are
\begin{eqnarray}
j_{\mu R}^3&=& { 1\over 2} \{
{\overline  {{u^m}_R}} \gamma_\mu  {{u^m}_R}  
 +{\overline  {{c^m}_R}} \gamma_\mu   {{c^m}_R} \nonumber \\  
 &+& ( \cos \theta_{tR})^2  {\overline  {t^m}_R } \gamma_\mu 
           {t^m}_R 
 +  ( \sin \theta_{tR})^2 
 {\overline  {T^m}_R } \gamma_\mu   
   {T^m}_R \nonumber\\
 &+& \sin \theta_{tR}   \cos \theta_{tR}   {\overline  {T^m}_R } \gamma_\mu   
   {t^m}_R  +(h.c.)\nonumber \\ 
 &-&  {\overline {d^m}_{Rj} } \gamma_\mu {d^m}_{Rj} \}
   ,
\end{eqnarray}
in both currents a sum over $j$ running from $1$ to $3$ is implied.
Because $\theta_{tR} \to {\pi \over 2}$ as $ M_T \to 0$,
there is a large mixing for
 $T^m_R$ 
in the right-handed charged current 
$j_{\mu R}^-$ and the right-handed neutral current $j_{\mu R}^3$.

%%%%%%%%%%%%%%%%%%% summary %%%%%%%%%%%%%%%%%%%%%%%%%%%%%%%
\section{\bf Summary}

We study a mechanism for the generation of the top quark mass
and flavor mixing in the context of a Seesaw model for
quark masses.
When the mass parameter $M_T$ for the isosinglet quark is 
smaller than the symmetry breaking scale of $SU(2)_R$, the lightest
of the two eigenstates is at the breaking scale of $SU(2)_L$
and can be identified with the top quark, the heaviest one
will be found at the symmetry breaking scale of $SU(2)_R$.
We also study the mass hierarchy of ordinary quarks by including the
flavor mixing. The stability of the
light quark masses against flavor mixing is explicitly shown.
The dependence on the strength of the Yukawa couplings is nontrivial
and we have given a simple geometrical interpretation for it.
The singlet-doublet mixing is suppressed
except in the case of the mixing between the right-handed top 
quark and its singlet partner.
This large mixing angle
% between righthanded top quark and its
%singlet partner 
is a characteristic of our model and may be
checked in the top quark production experiments.\\
%%%%%%%%%%%%%%%%%% acknowlegement %%%%%%%%%%%%%%%%%%%%
\section*{\bf Acknowledgments} 

We  would like to thank T. Muta and G. C. Branco for helpful discussions
 and A. Sugamoto 
for a suggestion.
We would like to thank Y. Koide for  conversations about his 
work and results. 
 This work is supported by the
Grant-in-Aid for Joint Internatinal Scientific Research
($\sharp 08044089$)
 from the Ministry of
Education, Science and Culture, Japan, and JSPS 
which made exchange programs of M.N.R. and T.M. 
 possible.

%%%%%%%%%%%%%%%%%%% Note Added %%%%%%%%%%%%%%%%%%
\section*{\bf Note Added}
After completing our work, we have received a
paper by Y. Koide and H. Fusaoka \cite{KF} in which the
enhancement of the top quark mass in a Seesaw model is studied.
They found that the top quark mass is enhanced to the
electroweak symmetry breaking scale at the
singular point ($b_f=-1/3$) of their singlet quark mass matrix
(democratic matrix + $b_f\times$ unit matrix). 
Their singular
point would correspond to $M_T=0$ in our case.
Ignoring flavor mixing, the mass scale we obtained for the 
fourth lightest up type quark 
together with the mass formula for the top quark coincides with
their result at the singular point.
Still, our approach differs considerably from theirs and
the mass formulae and the flavor mixing we obtain for the light quarks
are quite different. We attribute the mass hierarchy of the
five light quarks to the mass hierarchy of the five heavy 
singlet quark mass parameters and assume that the Yukawa couplings
among the singlet and doublet quarks do not depend significantly on the
flavour  whilst they start from a specific type of the singlet quark 
mass matrix and introduce flavor dependent Yukawa couplings.
As a result our formulae for the five light quarks cannot be 
directly translated into their framework. Futhermore their model
gives constraints on the CKM matrix unlike ours where we cannot make
definite predictions.
%%%%%%%%%%%%%%%%%%%% bibliography %%%%%%%%%%%%%%%%%%%%

%%%%%%%%%%%%%%%%%%%%%%%%%%%%%%%%%%%%%%%%%%%%%%%%%%%%
\appendix
\section*{\bf Appendix}
We prove the decomposition of $y_{L(R)}$ into a unitary matrix 
and a
triangular matrix.
Let us write a rank $3$ matrix $y$ with three complex vectors in
$C^3$,
\begin{equation}
y=( \vecg_1,\vecg_2,\vecg_3),
\end{equation}
where $y$ is  the singlet-doublet Yukawa coupling $y_{L(R)}$
in our case.
Choose three orthonormal vectors $U=(\vecu2_1,\vecu2_2,\vecu2_3)$ 
with $ \vecu2_3={\vecg_3 \over |\vecg_3|}$.
Multiplying $U^\dagger$ on the left hand side of $y$,
we obtain
\begin{equation}
U^\dagger y =
\left[ \begin{array}{ccc}\alpha_{11} & \alpha_{12}  & 0 \\
                             \alpha_{21} & \alpha_{22} &0 \\
                             \alpha_{31}  & \alpha_{32} &|\vecg_3|
  \end{array}
    \right],
\end{equation}
where $\alpha_{ij}=\vecu2_i^\dagger\cdot \vecg_j$.
Then we  define two vectors in  $C^2$,
\begin{eqnarray}
\valpha_1&=& \left[ \begin{array}{c} \alpha_{11} \\
                                               \alpha_{21}
\end{array} \right],\nonumber\\
\valpha_2&=&  \left[ \begin{array}{c} \alpha_{12} \\
                                               \alpha_{22}
                           \end{array} \right].
\end{eqnarray}
We also define two orthonormal vectors $\vecv2_1$ and $\vecv2_2$ with 
$\vecv2_2={\valpha_2\over |\valpha_2|}$ 
in $C^2$. With these two vectors, we can form another unitary
matrix $V$,
\begin{equation}
V= \left[ \begin{array}{ccc} \vecv2_1&\vecv2_2&\mbox{\boldmath $0$}\\
                        0 &  0  & 1
    \end{array} \right].
\end{equation}
By multiplying $V^\dagger$ to $U^\dagger y$,
we finaly obtain a triangular form,
\begin{equation}
    V^\dagger U^\dagger y =                  
\left[ \begin{array}
{ccc}   \beta_{11}  & 0 & 0 \\
        \beta_{21} & |\valpha_2| & 0 \\
        \alpha_{31} &\alpha_{32} &|\vecg_3| \end{array}
    \right],
\end{equation}
where $\beta_{ij}=\vecv2_i^\dagger\cdot\valpha_j$.
Then $y$ is written as the product of a unitary matrix and a triangular
matrix.
%%%%%%%%%%%%%%%%%%%%%%%%%%%%%%%%%%%%%%%%%%
\end{document}